\documentclass{eas}
\usepackage{graphicx}
\begin{document}

\title{Anisotropic Outflows and IGM Enrichment} 
\author{Hugo Martel}
\address{D\'epartement de physique, de g\'enie physique et
d'optique, Universit\'e Laval, Qu\'ebec, Qc, G1K 7P4, Canada}
\author{Matthew Pieri}
\sameaddress{1}
\author{C\'edric Grenon}
\sameaddress{1}
\begin{abstract}

We have designed an analytical model for the evolution
of anisotropic galactic outflows. These
outflows follow the path of least resistance, and thus
travel preferentially into low-density regions, away from cosmological
structures where galaxies form. 
We show that
aniso\-tropic outflows can significantly enrich low-density
systems with metals. 
\end{abstract}
\maketitle
\section{Introduction}

Galactic outflows play an important role in the evolution of galaxies
and the IGM. Supernova explosions in galaxies
create galactic winds, which deposit energy and metal-enriched gas into the
IGM. Simulations of explosions in a single
object reveal that outflows tend
to be highly anisotropic, with the energy and metal-enriched
gas being channeled along the direction of least resistance
(Martel \& Shapiro 2001).
Several observations also support the existence of anisotropic
outflows both directly
(e.g. Veilleux \& Rupke 2002) and indirectly (Pieri \& Haehnelt 2004;
Pieri, Schaye, \& Aguirre 2006).

\begin{figure}
\hskip0.6in
\includegraphics[width=3.3in,angle=90]{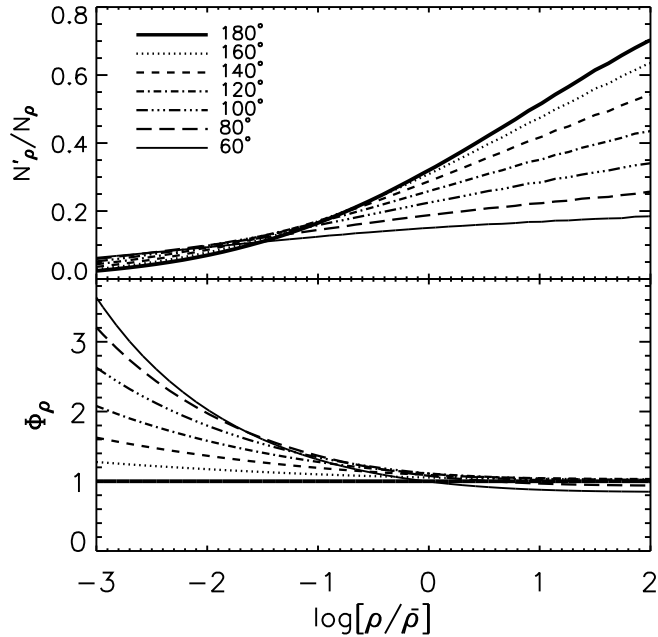}
\vskip-0.1in
\caption{{\it Top}: fraction of enriched grid points 
$N^\prime_\rho/N_\rho$ in the simulation volume at $z=2$ 
as a function of IGM overdensity $\rho/\bar\rho$
for varying opening angle. 
{\it Bottom}: number of enriched points below an overdensity threshold,
relative to the isotropic case.}
\end{figure}

We represent outflows as two spherical cones traveling in
opposite directions, along a path of least resistance, and
with an opening angle $\alpha$ that can vary from
$180^\circ$ (isotropic outflows) to $\sim20^\circ$. 
We describe the expansion of the outflows using the formalism of
Tegmark, Silk, \& Evrard (1993), modified for anisotropy.
To implement this outflow model in a cosmological simulation,
we use the Monte Carlo method of
Scannapieco \& Broadhurst (2001). We consider a comoving cubic volume of 
size $12h^{-1}{\rm Mpc}$, with periodic boundary conditions, in
a $\Lambda$CDM universe. 

\section{Results}

To investigate the nature of the regions enriched in metals by outflows,
we calculate the gas density and metallicity inside the computational
volume, on a $512^3$ grid. 
The top panel of Figure~1 shows the number of grid points $N^\prime_\rho$
enriched at a given overdensity $\rho$, as a fraction of the total number 
of grid points $N_\rho$ at that density,
for a range of different opening angles; this is 
effectively the probability 
of enriching a systems of a given density. The 
impact of galactic outflows on overdense systems is dramatically reduced for 
increasingly anisotropic outflows, while 
the probability of enriching low density systems 
increases.
The bottom panel shows $\Phi_\rho = N^\prime_{\rho^\prime<\rho}/
N^\prime_{\rho^\prime<\rho,180^\circ}$,
the number of enriched grid points with a overdensity 
below a given threshold $\rho$,
relative to the isotropic case.
This highlights the significant impact on 
the enrichment of underdense systems by anisotropic outflows. 
This can lead to an increase in the enriched volume of underdense systems 
of $10\%$ (where $\alpha=100^\circ- 120^\circ$) and an increase of $40\%$
in systems below $\rho/{\bar \rho}=0.1$ (where $\alpha=80^\circ- 100^\circ$) 
compared to isotropic outflows. 
 

\end{document}